# All-optical continuous tuning of phase-change plasmonic metasurfaces for multispectral thermal imaging

Matthew N. Julian[1,2], Calum Williams[3], Stephen Borg[4], Scott Bartram[4], and Hyun Jung Kim[2*]

[1] *Charles L. Brown Department of Electrical and Computer Engineering, University of Virginia, Charlottesville, VA, 22904, USA*
[2] *National Institute of Aerospace, Hampton, VA, 23666, USA*
[3] *Department of Physics, Cavendish Laboratory, University of Cambridge, Cambridge, CB3 0HE, UK*
[4] *NASA Langley Research Center, Hampton, VA, 23666, USA*

[*]*Corresponding author – hyunjung.kim@nasa.gov*

## Abstract

Actively tunable, narrowband spectral filtering across arbitrary optical wavebands is highly desirable in a plethora of applications, from chemical sensing, hyperspectral imaging to infrared astronomy. Yet, the ability to actively reconfigure the optical properties of a solid-state narrowband filter remains elusive. Existing solutions require either moving parts, have slow response times or provide limited spectral coverage. Here, we demonstrate a continuously tunable, spectrally-agnostic, all-solid-state, narrowband phase-change metasurface filter based on a GeSbTe (GST)-embedded plasmonic nanohole array. The passband of the presented tunable filter is ~74 nm with ~70% transmittance and operates across 3–5 μm – the thermal imaging waveband. Continuous, reconfigurable tuning is achieved by exploiting intermediate GST phases via optical switching with a single nanosecond laser pulse and material stability is verified through multiple switching cycles. We further demonstrate multispectral thermal imaging in the mid-wave infrared using our phase-change metasurfaces. Our results pave the way for highly functional, reduced power, compact hyperspectral imaging systems and optical filters.

**Word count:** *~3,200 (excluding Abstract, Methods, References and Figure captions)*
**Figure count:** *5*

# Introduction

Narrowband spectral filtering is integral in applications ranging from chemical spectroscopy, hyperspectral imaging to infrared astronomy [1,2]. For all major applications, multilayer interference (dichroic) filters offer unrivalled optical performance characteristics, yet unfortunately are passive. For tunable optical properties—critical for probing more than single wavelength and wide waveband operation—tunability is somewhat mimicked using motorized filter wheels containing several narrowband filters, or through mechanical tuning mechanisms [3]. Alternative frameworks for active tuning include Fabry-Perot-based micro-electro-mechanical systems [4-7], liquid crystal tunable filters [8,9] and acousto-optical tunable filters [10,11]. Nevertheless, all approaches suffer from inherent limitations, such as: having moving parts; being complex / expensive to manufacture; offering slow response times, and; providing limited spectral bandwidth / resolution. For compact, fast switching, narrowband spectral filtering across wavebands, no single solution currently exists.

In recent years, plasmonic nanostructures and metasurfaces have been proposed as potential solutions for tunable / reconfigurable filtering [12-18]. Through design, their optical response can be tailored to specific wavebands and thus specific applications. To date, these nanophotonic-inspired designs have primarily been passive; that is, their spectral response fixed post-manufacture. In contrast, tunable nanophotonic approaches, such as active metasurfaces, are capable of 'on-the-fly' dynamically tunable operation and have recently been demonstrated to show reconfigurable spectral filtering, thermal emission, beam steering, and tunable metalenses [12,14,16,17,19-26]. A number of different approaches have been utilized, including: thermally/electrically tunable structures based on $VO_2$ phase changes [22,23]; multi-quantum well (MQW) structures [24,27-29]; liquid-crystals [9, 30-32]; MEMS [4-7]; and epsilon-near-zero (ENZ) materials [33-35]. Nevertheless, translating these designs into multiple spectral bands is challenging. ENZ structures, for example, require operation at their respective ENZ wavelength, which inherently limits their operational spectrum. $VO_2$ has seen widespread utility as a phase-change material (PCM) in various spectral wavebands [22,23] but is not sufficiently transparent in the metallic state to be used for highly transmissive, real-world applications such as imaging or remote sensing, and is primarily suited for applications involving the modulation of a device's reflective or absorptive properties. Liquid crystals suffer from a similar drawback; their organic nature results in strong vibrational absorption bands in the mid-wave infrared (MWIR), limiting their utility across multiple wavebands. Additionally, liquid crystal-based nanophotonic color filters are generally limited in their tunability, bandwidth, efficiency, and are thus often utilized in reflective color generation [9, 30-32]. MQW structures have been demonstrated in an array of tunable photonic applications [24,27-29]. However, MQWs rely on metallic contacts to tune their effective refractive index (optical response), hence are more suited to reflective /absorptive applications, and are also not generally suited for the visible or near-IR spectrum. Tunable transmission filters using graphene plasmonic ribbons have recently been demonstrated [36]. However, a wide bandwidth, low transmission efficiency (~10%) and reduced long term chemical stability makes them equally unsuitable for the majority of real-world applications. Additionally, the reliance on the plasmonic resonance of graphene restricts these designs to IR/THz operation.

PCMs based on transition metal chalcogenide alloys such as GeSbTe (GST) are largely transparent across various spectral wavebands and exhibit significantly large, reversible refractive index modulation upon crystallization, making them ideal candidates for use in such spectrally robust filter designs [37]. Such 'exotic' optical materials have recently been utilized as the tunable medium for absorptive color filters [38-41], thermal emission switches [19,20], MWIR absorptive pixel arrays [42], and tunable metalenses [25,26]. GST has a dielectric permittivity of

$\varepsilon_r$~9.0 and $\varepsilon_r$~25.0 in its amorphous and crystalline states, respectively. Hence, GST provides an attractive tuning mechanism for resonance-based devices whereby the surrounding index strongly controls the spectral position of the resonance. Nonetheless, GST switching operation is generally considered binary in nature; operating in either its amorphous (a-GST) or crystalline (c-GST) state, subsequently limiting prospective devices to dual-modal operation as opposed to fully continuous. Theoretical studies have proposed absorptive devices based on partial crystallizations of GST, but experimental devices have not yet been realized [38]. Moreover, c-GST exhibits a non-negligible increase in its extinction coefficient compared to a-GST, which has thus far limited its use in transmissive applications. Other comparable PCMs, such as GeSbSeTe (GSST), have recently been proposed as low-loss alternatives to GST but are fundamentally limited by slow (> ms) switching times compared to the ns/ps response of GST [43].

In this work, we present for the first time, phase-change tunable metasurfaces based on GST-embedded plasmonic nanohole arrays (PNAs) with high transmission efficiency, narrowband performance, a continuous tuning range, and reversible operation. Reconfigurability is achieved through the continuum of partial-GST crystallinities using single nanosecond laser pulse induced phase transformation. These partial crystallinities—previously unexploited experimentally—exhibit repeatable, stable behaviour across a number of phase change cycles. Our results show excellent optical performance characteristics: high transmittance (~70%) on resonance, narrow bandwidth (~74 nm) within the 3–5 μm (MWIR) waveband, and near perfect reflection off-resonance. We show real-world applicability through multispectral thermal imaging demonstrations by integrating our metasurfaces with a MWIR camera. Our device design framework can be tailored to operate across any optical waveband through phase-change alloy selection. The results presented represent a significant step toward robust, tunable spectral filters, with applications in compact, fast-switching, and all-solid-state tunable filter systems for imaging in a wide variety of fields, from astronomy to remote Earth sensing.

# Results

## GST thin-film morphology

Initially, to verify thin-film quality, GST-225 films were deposited onto CaF$_2$ substrates via RF magnetron sputtering from a GST-225 target. Detailed descriptions of the GST deposition and characterization are given in the *Methods* Section, *Supplementary Note 1, Supplementary Figures S1-S3*. The stoichiometric ratios of deposited films were confirmed via direct current plasma atomic emission spectrometry (DCP-AES) measurements. The crystallinity and film thicknesses were determined via X-ray diffraction (XRD) measurements and SEM of cleaved samples, respectively. The complex refractive indices of the films were characterized using IR-variable angle spectroscopic ellipsometry (J.A. Woollam, Co.) in the spectral range from 1–10 μm. Results of both the XRD and ellipsometry measurements of as-deposited GST-225 are shown in **Fig. 1**—for a-GST and both hexagonal-centered cubic (HCC) and face-centered cubic (FCC) c-GST. The XRD data (**Fig. 1**a-c) shows a clear transition from amorphous (**Fig. 1**a) to FCC and HCP states (**Fig. 1**b,c). The two crystal states share the same main peak at $2\theta = 30°$, but show a stark difference in the magnitudes of the peaks centered around 42° and 26°, indicating the different GST unit cell configurations. The transition from a-GST to c-GST shows the expected large refractive index shift $\Delta n' = 2.0$, with small increase in extinction coefficient as wavelength increases (**Fig. 1**d,e) [44]. As can be seen in the characterization results in **Fig. 1**f, the as-deposited films show very low surface roughness and high uniformity in thickness.

## Phase-change plasmonic metasurface

Our GST-based PNA MWIR metasurface concept is shown in **Fig. 2**. The extraordinary optical transmission (EOT) response, thus local field enhancement, of PNAs strongly depends on the surrounding index [45-48]. Typically, this is exploited by modulating the index above or below the hole array [9,30-32,36]. However, the resonance is particularly sensitive to the index inside of the individual holes—a fact that has been largely unexploited experimentally [48]. Here, the GST-PNA metasurface response is therefore dependent on the crystallization state of GST, which is subsequently tuned (continuously) through differing partial crystallizations, using a ns laser pulse (**Fig. 2**a,b). This operating principle is summarized in **Fig. 2**c, whereby an increasingly large pump energy incident upon the active area continuously adjusts the GST crystallinity, changing its index, hence spectrally shifting the surface plasmon resonance (SPR)-mode and changing the resultant transmission response of the device. A 'reset pulse' (high energy) is then used to revert the device back to a-GST, thus allowing for repeatable operation. Using the experimentally measured complex refractive index of p-GST (**Fig. 1**d,e), electromagnetic simulations (Lumerical FDTD [49]) of the GST-PNA device—shown schematically in **Fig. 2**d,i—were performed (**Fig. 2**d) in order to optimize geometric parameters for narrowband optical performance across 3-5 μm. Ag is chosen due to its superior optical properties across the MWIR in comparison to other noble metals (e.g. Al, Au). At resonance, the large field enhancement provided by the SPR (**Fig. 2**d, ii) partially mitigates the non-zero c-GST extinction coefficient, leading to a larger-than-expected transmission efficiency. The SPR-origin of the enhancement phenomenon—consistent with EOT—was confirmed by performing simulations with non-plasmonic metal films (*Supplementary Figure S4*). The hole diameter of the hexagonal array was optimized via sweeping the ratio of diameter to period at a given period; a ratio of 0.4*period was found to be optimal with 1800 nm array period and 60nm device thickness (*Supplementary Figure S5*). The effect of array periodicity and embedded refractive index GST/Ag PNAs (with 720 nm hole diameter) is presented in **Fig 2**. d (iii-iv). As expected, a linear dependence on both quantities is observed, and the simulated

FWHM is consistently ~80nm throughout the simulated range (with a slight increase at longer resonance wavelengths), with perfect reflection shown at the off-resonance (*Supplementary Figure S5*). The simulations performed show a tuning range of 2.9µm–4.7µm, however by further increasing/decreasing the period of the nanohole array, this range can be extended to include the LWIR/NIR.

The metasurface devices were designed for an amorphous-state (a-GST) resonance at 3.0 µm, which corresponds to a period of 1800 nm (hole diameter 720 nm). A SEM micrograph of one of the GST-PNA devices is shown in **Fig. 3**a. 60 nm of Ag was deposited via magnetron sputtering onto $CaF_2$, patterned via direct-write photolithography and dry etched to generate the nanohole array, the total patterned device area was 15 mm x 15 mm (see *Methods* section for further details). 60 nm of GST was then conformally deposited such that the holes were filled with GST. The presence of conformal GST atop the Ag film has a negligible effect on device performance. Lastly, an encapsulation layer of $ZnS:SiO_2$ was deposited in order to prevent oxidation of the Ag/GST as well as any partial volatilization that may arise as a result of switching. This layer is fully transparent to across the visible and MWIR wavebands. Full details of device characterization comparing the capped/uncapped devices are provided in the *Supplementary Note 2* and *Supplementary Figure S6*.

Transmission-mode Fourier Transform IR-spectroscopy (FTIR) was used to optically characterize the GST-PNA metasurfaces. A laser pulse induces crystallization of the as-fabricated a-GST devices, it is then re-characterized in its crystalline state (c-GST). The resultant transmission response due to the respective GST-state is shown in **Fig. 3**b. FTIR data is shown for both the HCP and FCC phases of c-GST. Samples were subsequently heated to melting temperature by applying a single 160 $mJ/cm^2$ 100 ns laser pulse to return to the amorphous phase to be re-characterized. A total of ten complete phase-change switching cycles was performed, while recording resonance peak position (**Fig. 3**c), to demonstrate the stability of the filtering response. The devices show high stability across all cycles in both the amorphous and crystalline phases and the shape of the spectral response in each phase is unchanged (**Fig. 3**d), which is a result of the addition of the capping layer. The devices show a transmission efficiency of ~70% in both crystalline phases, which is among the highest transmission for such a structure reported in the literature, irrespective of spectral band. A FWHM of 74 nm in the amorphous phase and 100 nm in the crystalline phase, is in good agreement with simulation results. Moreover, the peak centre wavelengths lie at ~2.9 µm in the amorphous state and 3.35 µm in the crystalline state, which is also within good agreement to the simulated values of 3.0 µm and 3.5 µm respectively. This small deviation is likely the result of either batch-to-batch variation in the refractive index of the deposited GST, microfabrication tolerances, and/or as a result of the addition of the capping layer.

## All-optical continuous tuning

For many applications, active optical tuning is highly desirable, i.e. the passband centre wavelength can be spectrally shifted in real-time. This is traditionally achieved through optical, electrical or thermal stimulus, but an all solid-state solution has been challenging. We implement an all-optical approach, demonstrating continuous spectral tuning of a ~8.4 mm$^2$ device with a single ns laser pulse. The setup is shown schematically in **Fig. 4**a: A 532 nm laser (100 ns) is used to rapidly switch the GST crystallinity, thus output optical response. The pulse energy, with 7mm beam diameter, was increased linearly from 2 - 250 mJ, in order to study the effect of incident pulse energy on the GST crystallinity. A laser spot size of 8.4mm was chosen due to the 6.5 mm measurement area probed by the FTIR system. By varying the fluence from 10–60 mJ/cm$^2$, intermediate phases of GST were generated, with full crystallization being achieved at ~60 mJ/cm$^2$ fluence. Here, three unique p-GST states were formed, with refractive index values ranging from ~3.0–3.75, and $\Delta n$ ~ 0.25 between each state. By implementing fine control of the pulse energy (e.g. with a polarizer and half-wave plate) a virtual continuum of p-GST states can be generated (i.e. $\Delta n$ between adjacent states can be minimized and the number of states increases). In order to return to the amorphous phase, it was found that a fluence of 160 mJ/cm$^2$ was required (i.e. 'reset pulse'). Ellipsometry data and FTIR characterization of the optically tuned GST films (a-GST, c-GST, and two p-GST states) are shown in **Fig. 4**b-d. The transmission response in each state is approximately constant despite the small increase in extinction coefficient with increasing GST crystallinity. This is due to the increased field enhancement observed in PNA devices filled with a high-index dielectric medium [39]. The peak centre wavelength as a function of laser energy density (**Fig. 4**e) exhibits linear behaviour, which is in strong agreement with simulations. Moreover, the a-GST devices—before switching and after laser-induced return (i.e. 'reset pulse' to the amorphous state)—show identical spectral responses, demonstrating stability in the optical tuning mechanism. As an additional benefit, due to the conformal deposition of the GST films, there is essentially zero absorption in the Ag film as a result of the shallow (~5 nm) absorption depth of GST at 532 nm [50]. This will enable long device lifetimes and stable operation across many switching cycles. To the best of our knowledge, this is the first time tunable operation across the MWIR with high transmission efficiency (~70%) and narrowband filtering (~70nm FWHM, Q-factor ~45) has been achieved.

## Actively tunable thermal imaging

The thermal radiation from an object is described through its blackbody emission curve, whereby the cooler an object the longer the peak wavelength of its emission curve. A contact hotplate for example, with only a relatively small temperature change from 320K (47°C) to 480K (213°C) exhibits significantly different spectral emissive power responses (**Fig. 5**a). To demonstrate real-world imaging applicability, we use our tunable GST-PNA metasurface filters in combination with a contact hotplate (thermal source) and commercial FLIR MWIR camera (imager). A setup schematic is shown in **Fig. 5**b (image of setup shown in *Supplementary Figure S*7). Fixing the source temperature at ~480K, our metasurface filters provide tunable filtering operation across the region 2.91–3.41 μm (**Fig. 5**c) based on the switching between five different GST-crystalline states (from a-GST to p-GST to c-GST). Because the blackbody emissive power from a 480K source increases at the longer wavelengths in our targeted spectral region, we expect the tunable filter to provide an increasing transmitted intensity; this is clearly observed in **Fig. 5**c. A percentage comparison between expected and experimental transmittance, used to validate our response, is shown in *Supplementary Figure S8*. To further demonstrate thermal imaging applicability, in particular to verify transmission homogeneity across the filters through spatial variation of the thermal source, we fix a metal (Al) transmission mask (3mm thick) with NASA Insignia logo in front of the thermal source (**Fig. 5**d). By gradually increasing its temperature (from $T_o$ = 320K–480K) and optically tuning two separate fabricated GST-PNA filters to exhibit passband centre wavelengths of 2.91 µm and 3.41 µm, we are able to show selective MWIR imaging of the logo at two separate passbands simultaneously (**Fig. 5**e). The accompanying imaging video can be found in *Supplementary Movie M1*.

# Discussion

In summary, we have experimentally demonstrated continuously tunable narrowband GST-PNA based metasurface spectral filters operating across the MWIR by exploiting intermediary partial crystallinities of the phase-change material GST. We further show real-time multispectral thermal imaging across the MWIR waveband, by integrating our GST-PNA metasurfaces with conventional IR-camera. Our phase-change tunable metasurfaces show both the highest transmission efficiency (~70%) and narrowest bandwidth (~74nm) in relation to the wider tunable filter community—a direct result of the increased field enhancement, and mitigating the extinction losses of c-GST, arising from the large refractive index contrast of the embedded GST within the nanoholes [48] rather than underneath / above the holes, which results in a low-Q, low transmission, and limited tunability [9,31,36,51-53]. Through the utilization of p-GST states—a relatively unexplored concept experimentally—our continuously tunable phase-change metasurfaces represent a fundamentally new spectral filter, opening the door for both high efficiency, high-Q PNA-based metasurface photonic devices. In comparison to other approaches for reconfigurable / tunable filters [3,4,9,12,13,14,16,17,36], the devices here are able to operate continuously, across a spectral range of several microns (3–5 µm), along with possessing a simple, spectrally agnostic design that is translatable to any waveband. By taking advantage of the fine-tuning provided by optical switching, we have experimentally demonstrated the utility of partially-crystalline GST states in order to achieve continuous tuning across the spectral range of operation and shown the tuning is stable across a number of switching cycles.

From an applications standpoint, the PNA design is straightforward to manufacture and is capable of switching at ns speeds, making it attractive for systems with time resolution requirements. Through industry standard UV-lithographic and physical vapour deposition techniques, the design scheme for the 1-inch GST-PNA optical elements presented is easily scalable to larger areas. Interestingly, a virtual continuum of crystallinities, thus continuum of passbands, can be achieved by controlling the laser pulse energy. The large pulse energies used here are a result of the relatively long pulse width (100 ns) of the laser used, and the large area (~55 mm$^2$) being switched with a single pulse. Much smaller pulse energies can be used with a shorter pulse duration and/or by employing a raster scanning technique with a reduced beam size to increase the overall fluence [54].

The work presented represents a significant improvement in the state of the art of PNA and GST metasurface devices, as well as tunable optical filters. This proof of concept demonstration is achieved with a nanosecond pulse laser system; however, the general framework can be translated to a much more compact system for real world system integration (*Supplementary Note 5 and Supplementary Figure S9*) with switching implemented through an electrical stimulus (akin to rewritable optical storage media designs). Furthermore, the transmission efficiency may be further increased by employing an anti-reflection coating to each side of the device, or by reducing the extinction coefficient of the GST film. We stress that although GST is opaque in the visible, other phase-change alloys such as SbTe, GeTe, or a custom blend of chalcogenides, can be used for similar device operation in the visible spectrum [55,56], and thus the general design framework shown here is spectrally agnostic. As a result, we expect the promising findings presented here to be useful not only in fundamental tunable photonic device / metasurface research, but in a host of applications including hyperspectral imaging, remote sensing, and chemical spectroscopy.

# Methods

## Sample preparation and film deposition

Double-side polished 25.40 mm diameter (1-inch optics), 1.50 mm-thick calcium fluoride ($CaF_2$, Esco Optics, Inc) wafers were cleaned in acetone, iso-propanol, and deionized water sequentially, twice, before deposition. The $Ge_2Sb_2Te_5$ thin films were deposited via RF magnetron sputtering at a base pressure of 2.6 x $10^{-6}$ Torr and a deposition pressure of 7 mTorr (10 sccm Ar flow, research-grade, 99.9999% purity). A $Ge_2Sb_2Te_5$ target (14.3 wt% Ge, 23.8 wt% Sb, 61.9 wt% Te, Mitsubishi materials, Inc.) was sputtered; 50W RF power at room temperature and 10 rpm substrate rotation during deposition. 10 min pre-sputtering was conducted at 7mTorr with 50 sccm of Ar flow with the target shutter open. After 30 minutes of deposition, the growth rates of GST were measured via SEM (VEGA3, TESCAN) as 6.7 nm/min. Prior to XRD and ellipsometry characterization, the as-deposited amorphous phase GST film on $CaF_2$ wafer was annealed at 145°C (*Supporting Information 1*) and 172°C inside the sputtering chamber for 1 hr.

For the PNA device fabrication, 60 nm Ag film was deposited on $CaF_2$ substrates by using the RF sputtering at a base pressure of 6.6 x $10^{-6}$ Torr and a deposition pressure of 3 mTorr. The 14.7 Å/sec slow deposition rate is selected for the precise control of the 60 nm film thickness. ~800 nm thickness AZ1500, high resolution photoresist, was spin coated at 4000 rpm and soft baked at 100 °C for 60 s. Direct-write UV-laser lithography (DWL 66fs, Heidelberg Inc.) was used for nanohole patterning with 2-mm writing head. Post-development (AZ 300MIF developer), Ag thin films were dry-etched using inductively coupled plasma (ICP, TRION Tech.) equipment in $CF_4$ / Ar mixed-gases while maintaining 15 mTorr pressure, 500 W inductive power, and 150 V DCbias voltage. The GST film was conformally deposited after the nanohole patterned on Ag / $CaF_2$ sample. AZ100 remover was used as chemical stripper.

## Material characterization

*GST composition ratio*: The chemical composition of as-deposited GST thin films was determined by DCP-AES (Direct Current Plasma-Atomic Emission Spectroscopy, Luvak, Inc.). The composition was measured as 22 at% Ge, 23.5 at% Sb, and 54.5 at% Te, which is close to a nominal composition of bulk $Ge_{22.2}Sb_{22.2}Te_{55.6}$. For the DCP-AES test, 1 μm thick GST film was deposited on Si(100) substrate and fully dissolved in 1:1 ratio of hydrochloric acid (HCl) and hydrofluoric acid (HF) to collect and detect the individual Ge, Sb, and Te elements [57,58].

*Refractive index and extinction coefficient*: The complex refractive indices of the bare $CaF_2$ substrate and the $Ge_2Sb_2Te_5$ film on $CaF_2$ were measured over the wavelength range of 193 nm

-33 µm. The RC2® ellipsometry system (Model: DI, J.A. Woollam) used to measure the properties from 193nm-1690nm spectral range with 55° to 75° angles of incidence by 10° step and the Complete EASE v.6.33 used for the data analysis. IR-VASE (Infrared Variable-angle spectroscopic ellipsometry, J.A. Woollam) was used to measure the optical properties from 1.7-33 µm and transmission intensity data from 1.7-13.5µm at angles of incidence of 55° to 75°, increments of 10°. Data analysis was performed using WVASE v3.908 software.

## Numerical simulations

Finite Difference Time Domain (FDTD) simulations were performed using a commercial electromagnetic simulator (Lumerical FDTD Solutions [49]). 3D simulations for GST-PNA metasurface device geometry were performed using a plane wave (full spectral coverage), symmetric boundary conditions in the *x* and *y* dimensions, and perfectly matched layer boundaries in *z*. A 10 x 10 x 5 nm mesh size was used across the device itself. Simulations and parameter sweeps were allowed to converge for each iteration. Index and E-and-H-field profile monitors were placed at various positions along the z-axis in order to accurately monitor the simulation. A power monitor was placed in the far-field of the device in order to remove any near-field effects that may be present while collecting transmission data. Complex dispersive material models were used for Ag (Johnson and Christy model) and $SiO_2$ (material data), whereas GST thin-film index models were implemented using experimentally determined values (IR-ellipsometry).

## IR Imaging Demonstration

IR-images were captured using the setup schematically shown in **Fig. 5**b (photographs of the setup shown in *Supplementary Figure S7*). An IR camera (FLIR SC8300HD, 0.02 W/m² minimum pixel sensitivity) was used to record the images. A field stop was used in order to eliminate unwanted signal from polluting the measurement data. A NASA insignia logo was milled into a 15 cm x 15 cm x 3 mm aluminum plate and used in front of a contact hotplate which was used to mimic a blackbody thermal source. The hotplate was heated to 486K and allowed to stabilize for the measurements and images shown in the figure. For the video demonstration (*Supplementary Movie 1*), the hotplate/object was heated from 320K to 486K.

## Figures

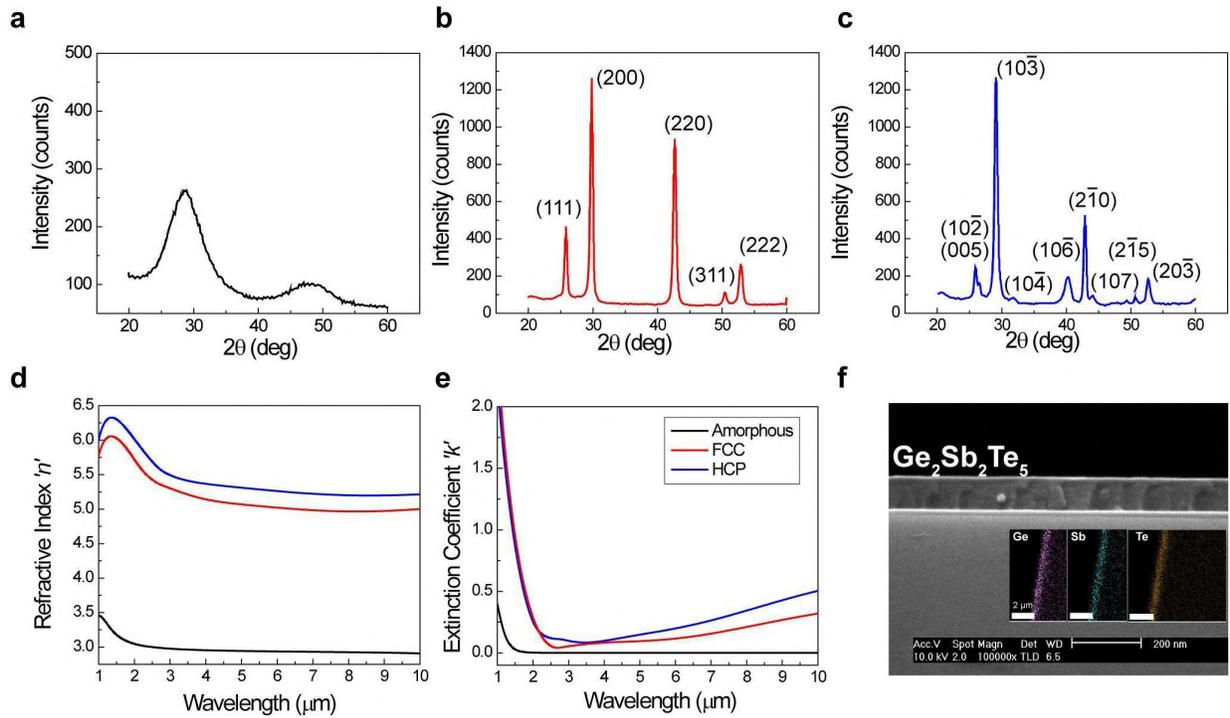

**Figure. 1. Thin-film GST characterization**

XRD (a-c) and ellipsometry (d, e) data for a-GST (black curves), FCC c-GST (red curves) and HCP c-GST (blue curves). The average Δ$n$ is ~2.0 between a-GST and c-GST, with HCP c-GST exhibiting slightly higher refractive index and extinction coefficient compared to FCC c-GST. (f) SEM cross section image of the GST film deposited on CaF$_2$. The inset shows a top-view energy-dispersive x-ray spectroscopy (EDS) mapping of the film edge, showing the clear presence of Ge, Sb, and Te species.

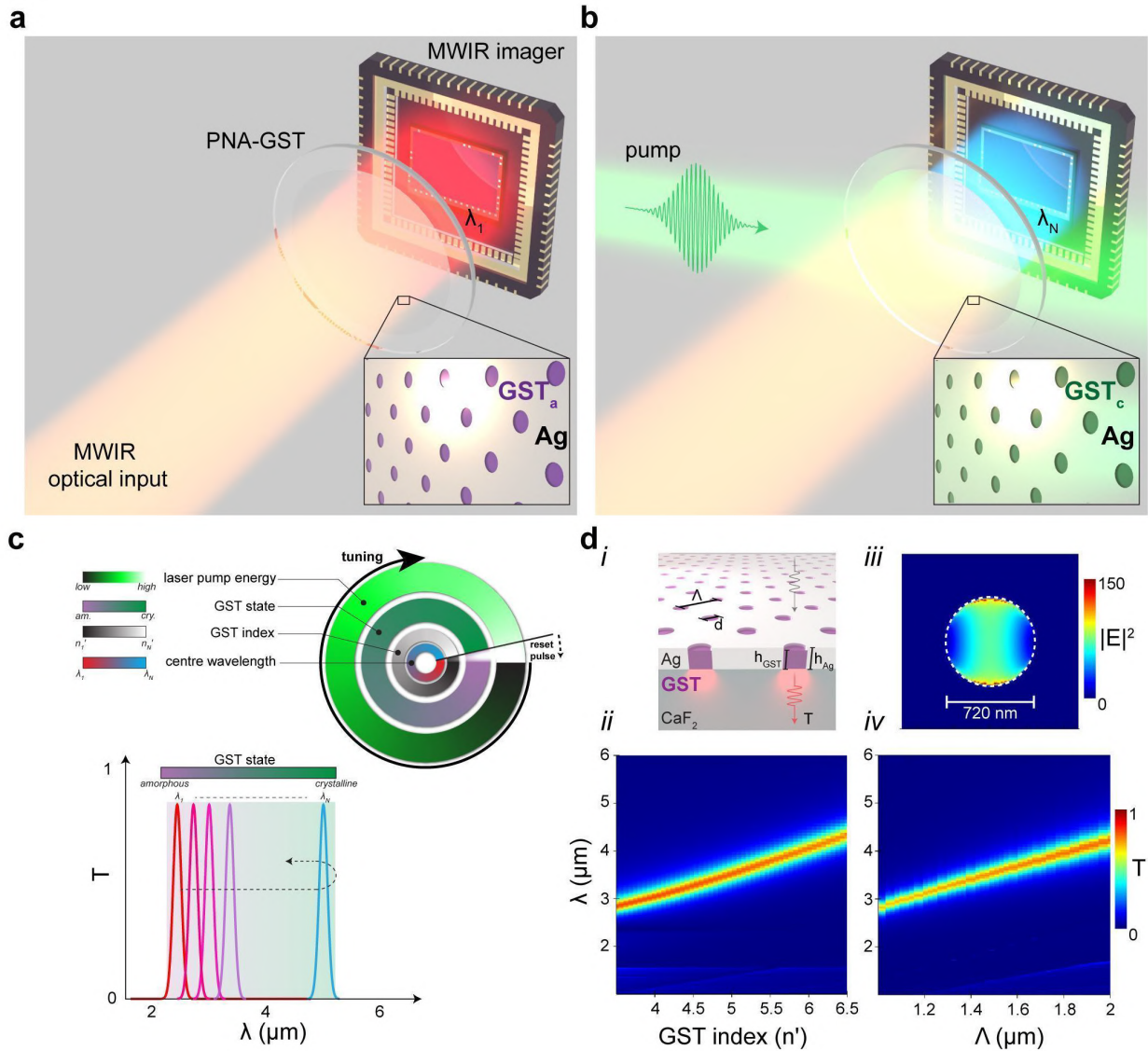

**Figure. 2. Tunable GST-plasmonic nanohole array metasurface for the MWIR waveband**

Device concept whereby the MWIR optical input is imaged through PNA-GST filer in its initial, amorphous state (a), with initial centre wavelength, $\lambda_1$. Through a laser pulse incident on the PNA-GST area, the GST crystallinity is modified (phase change) and the resultant transmission response (centre wavelength, with initial centre wavelength, $\lambda_N$) is spectrally shifted (b). This behaviour is summarized in (c) whereby the pump energy controls GST-state, which in turn changes its refractive index, hence spectrally shifts the centre wavelength from the resonant PNA. A 'reset pulse' returns the GST to its initial state, thus device to initial transmission centre wavelength. (d) FDTD simulations of a GST-embedded PNA (i). The simulated transmission response of the GST-PNA device as a function of GST refractive index (ii) and as a function of period (iv) for a film thickness of 60 nm and hole diameter of 0.4*period. E-field plot (iii) showing SPR-generated field enhancement on-resonance at the boundary between Ag /GST inside of the nanohole cavity.

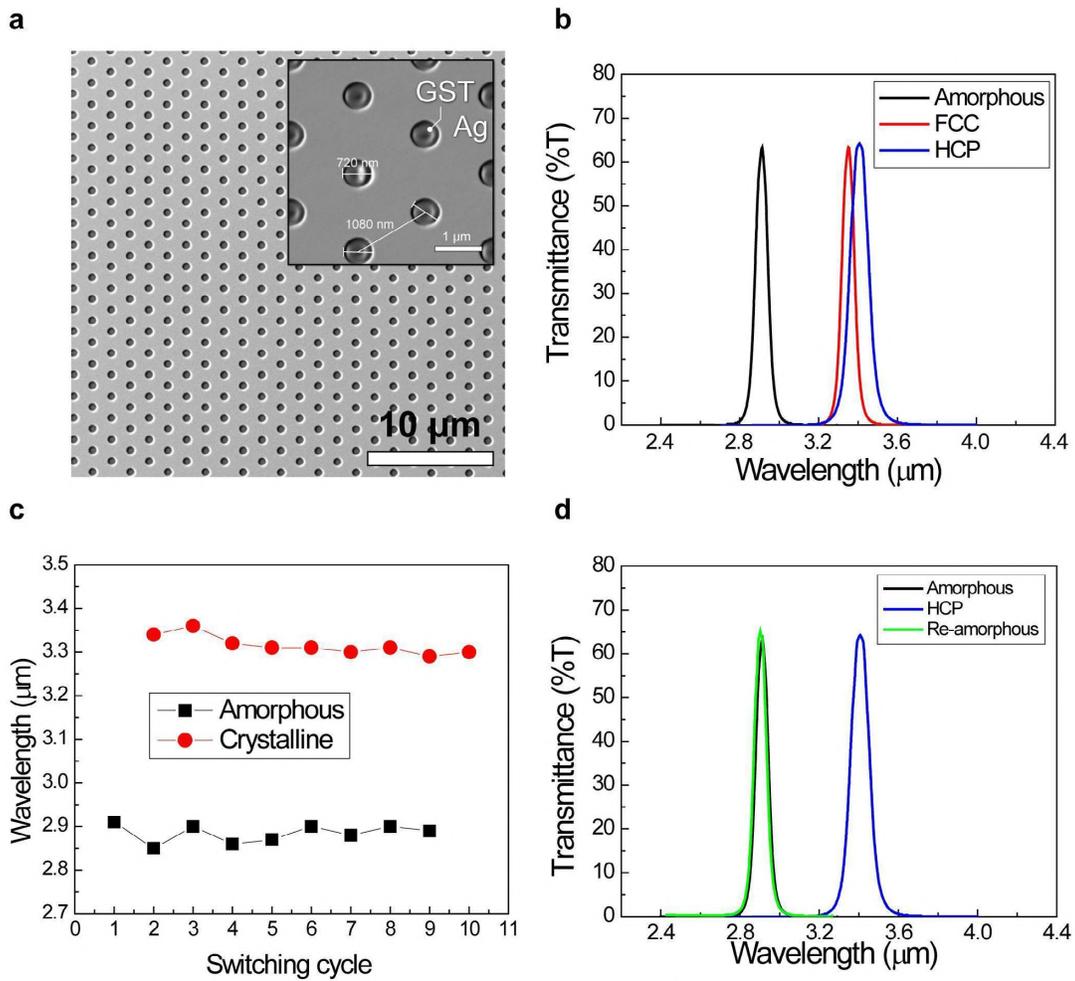

**Figure. 3. GST-PNA device tuning**

(a) SEM micrographs of the fabricated tunable GST-PNA metasurface device showing the full hexagonal array geometry and individual hole morphology (inset). The GST embedded within the Ag PNA can be seen. (b) FTIR (transmission) characterization of the fabricated PNA device showing ~70% transmission at the resonance and perfect reflection outside the resonance bandwidth. Stability in the spectral response was maintained across many switching cycles; shown through centre wavelength reproducibility (c) and spectral shape (d) consistency.

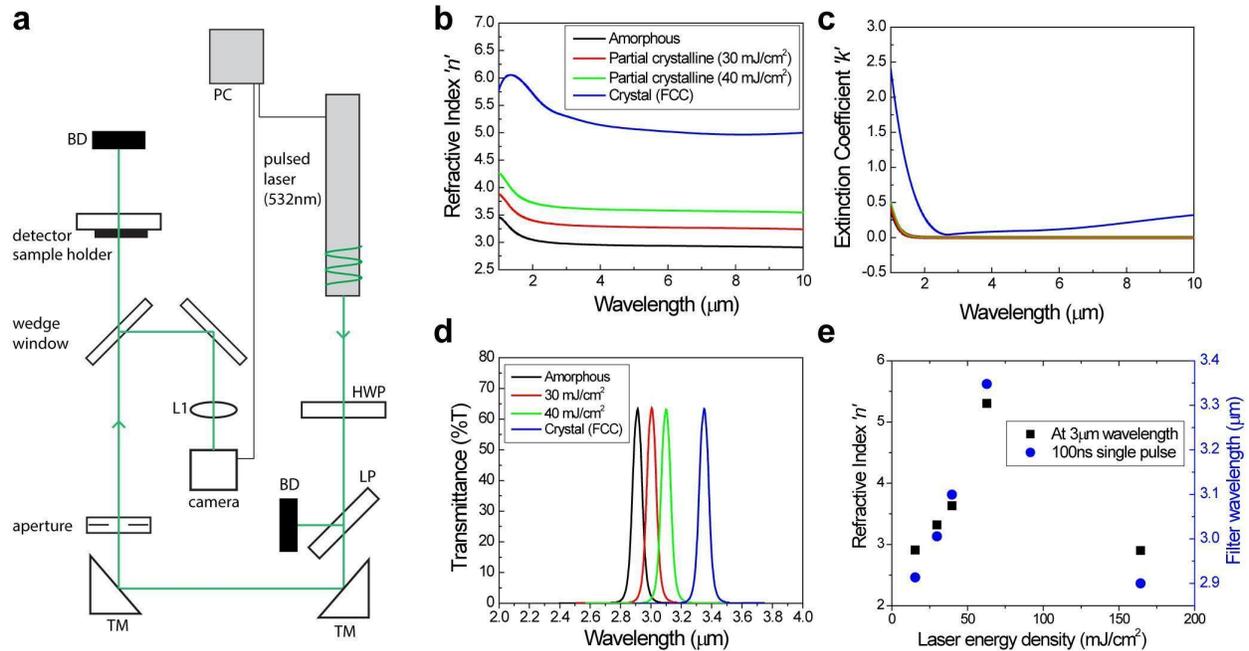

**Figure. 4. Optically-tuned GST-PNA metasurface devices**

(a) Setup used for the laser switching demonstration. Complex refractive index measurements (b, c) of the a-GST, p-GST, and c-GST films (tuned using the all-optical approach), along with the corresponding spectral response of the full GST-PNA metasurface device for each case (d). It can be seen in (e) that with increasing pulse energy, the crystallinity increases approximately linearly until c-GST is achieved. Further increasing the pulse energy allows for the return to a-GST. Upon returning to the amorphous state, the device exhibits nearly identical spectral response to the as-deposited amorphous phase device.

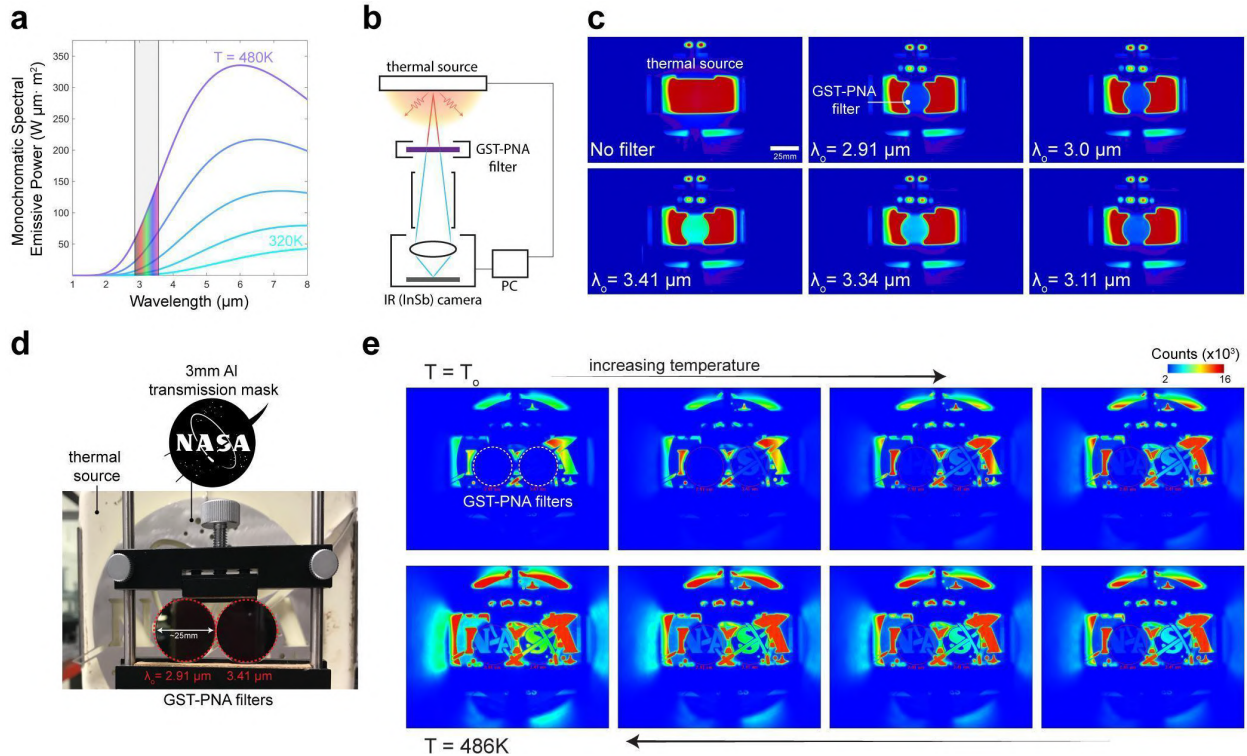

**Figure. 5. MWIR imaging using tunable GST-PNA metasurface filters**

(a) Blackbody thermal emission curves for varying temperature sources with overlaid spectral coverage of the tunable GST-PNA metasurface filters fabricated here. (b) Thermal imaging setup schematic for the results shown in (c,e); image of setup shown in *Supplementary Figure S7*. (c) MWIR imaging results at a fixed 486K hotplate temperature, as a function of varying GST-PNA filter states with varying passband centre wavelength, $\lambda_o$. (d) RGB image of the setup imaged in (e), which shows the IR image of the same scene, as the temperature of the hotplate is increased from 320K to 486K. The *left* and *right* filters are centered at 2.91 μm and 3.41 μm, respectively. Variable transmission response through the filters, and subsequent identification of the logo (spatially variant thermal profile), can be observed.

# Acknowledgements

M.J., S.B., S.B., and H.J.K acknowledge support from the NASA LaRC CIF/IRAD Program. C.W. acknowledges support from the Cancer Research UK Pioneer Award (C55962/A24669), Engineering and Physical Sciences Research Council (EP/R003599/1) and through the Wellcome Trust Interdisciplinary Fellowship. The authors appreciate Dr. Nina Hong of J.A. Woollam Inc. for assistance with the ellipsometry analysis and Mr. Joel Alexa at NASA LaRC with XRD analysis.

# Author Contributions

M. J., C.W. and H. J. K. conceived the idea and designed the device. C.W. and M. J. performed the simulations and design optimization. S. B. and H. J. K. performed GST film growth and characterization, device fabrication, and device characterization. M. J. and C. W. advised the device characterization. S. B. and H. J. K. carried out the optical switching measurements. M. J. and C. W. designed the imaging demonstration. H. J. K and S. B. carried out the imaging demonstration and analyzed the data. All authors wrote the manuscript.

# Competing interests

The authors declare no competing interests.

# Data availability

The data that supports the findings of this work will be made available online (open access) with DOI upon publication.

# SUPPLEMENTARY INFORMATION

All-optical continuous tuning of phase-change plasmonic metasurfaces for multispectral thermal imaging

---

- **Supplementary Note 1:** GST Deposition and characterization
- **Supplementary Note 2:** FDTD simulations
- **Supplementary Note 3:** Laser characterization setup
- **Supplementary Note 4:** IR imaging setup
- **Supplementary Note 5:** Active "Read/Write" System Concept -- Laboratory and Real World Systems

# Supplementary Note 1:
## GST Deposition and characterization

For the XRD and ellipsometry tests, the as-deposited amorphous phase GST film on $CaF_2$ wafer was annealed at 145°C (200°C temperature setting) and 172°C (240°C temperature setting) inside the sputtering chamber for an hour. The substrate temperature of the sputtering system was calibrated by using the thermocouple wafer on the graphite sample holder.

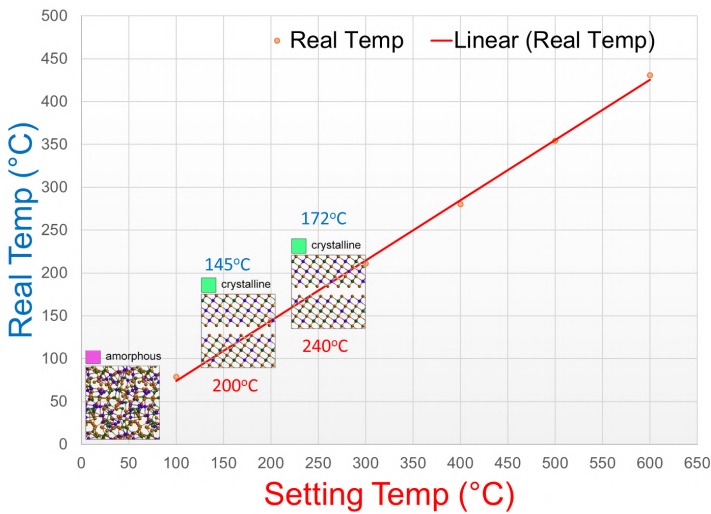
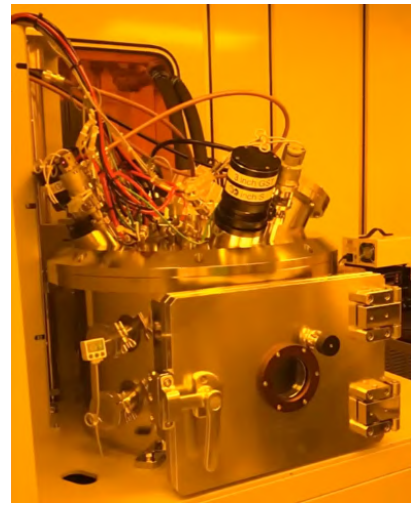

**Supplementary Figure S1**
*Substrate temperature calibration (left) of the sputtering chamber (photo, right) by using the thermocouple wafer and graphite sample holder.*

## GST Composition Ratio

| Sample | Ge | Sb | Te | Melting point (ºC) |
|---|---|---|---|---|
|  | Chemical composition as at% (atomic ratio) | | | |
| Lab grown specimen | 22 (2) | 23.5 (2) | 54.5 (5) | 611 |
| Stoichiometric composition | 22.2 | 22.2 | 55.6 | 615 |

**Supplementary Figure S2**
*Table showing the characteristics of the ideal and lab-grown GST-225 specimen. Samples show near-idealized values for atomic ratio %s as well as melting temperature.*

## GST film thickness change by phase change (annealing)

It is well known that GST will contract slightly when transitioning from amorphous to crystalline phases. These thickness changes were measured so that their effect of PNA device performance could be predicted and understood (effects are shown in the simulation results in Figure S4).

|       | GST film thickness (nm) | | | GST film thickness (-%) | | |
|-------|-------|-------|------|------|------|------|
| RT    | 200nm | 100nm | 25nm |      |      |      |
| 170°C | 189nm | 94.6nm | 24nm | 5.5% | 5.4% | 4% |
| 260°C | 187nm | 94.6nm | 22.7nm | 6.5% | 5.4% | 9.2% |
| 360°C | 185nm | 92.7nm | 22.7nm | 7.5% | 7.3% | 9.2% |

**Supplementary Figure S3**

*Summary of the change in GST film thickness as measured on a $CaF_2$ via SEM cross sectional analysis. The film contracts as it becomes more crystalline (i.e. increasing crystalline fractions of p-GST). Contractions of nearly 10% were observed.*

# Supplementary Note 2:
## FDTD Simulations and Device Characterization

Additional FDTD simulations were performed to understand the effect of various parameters of the PNA device performance. Film thickness, metal film material, (hole diameter)/(array period) factor, and reflection spectrum were all simulated in order to optimize the device as well as confirm the role of surface plasmon resonance (SPR) and lack of absorption in the device away from the resonance condition.

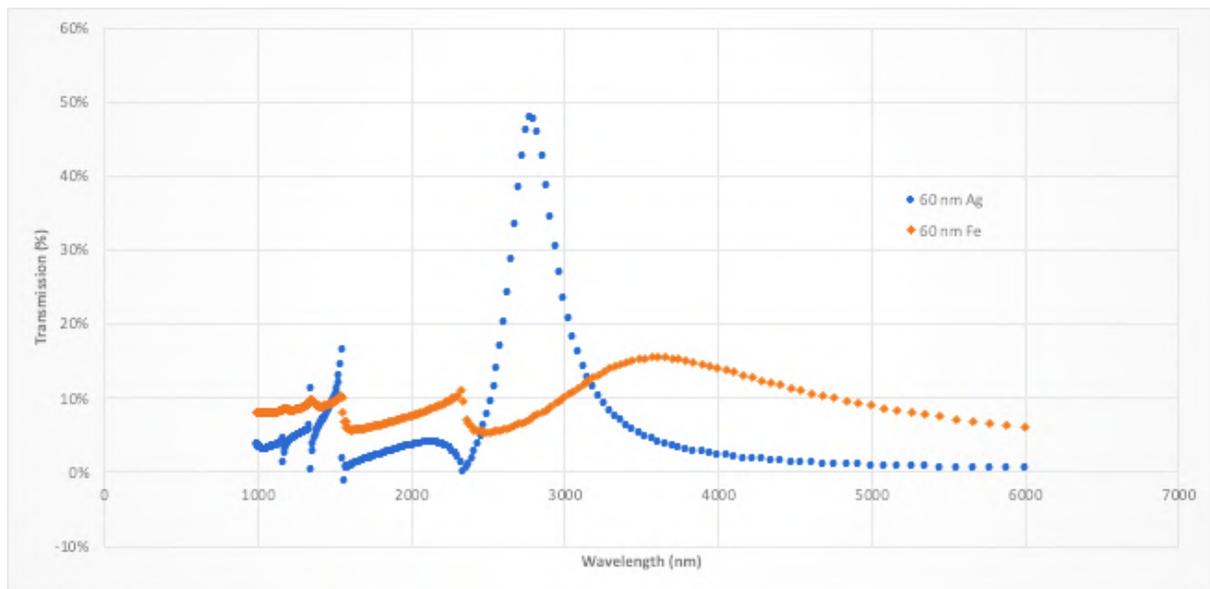

**Supplementary Figure S4**
*Comparison of an Ag metal film to an Fe metal film (t = 60 nm for both cases) with a-GST filling the holes in the metal films. The Fe film clearly lacks a SPR, as is expected for a non-plasmonic film such as Fe.*

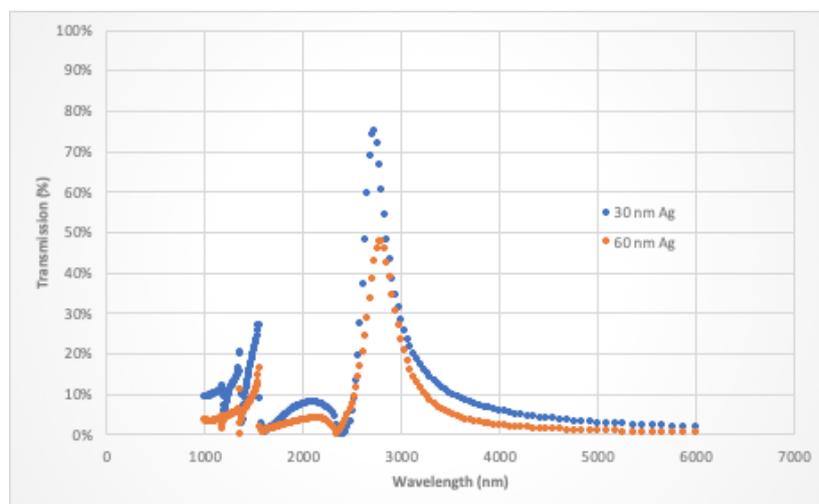

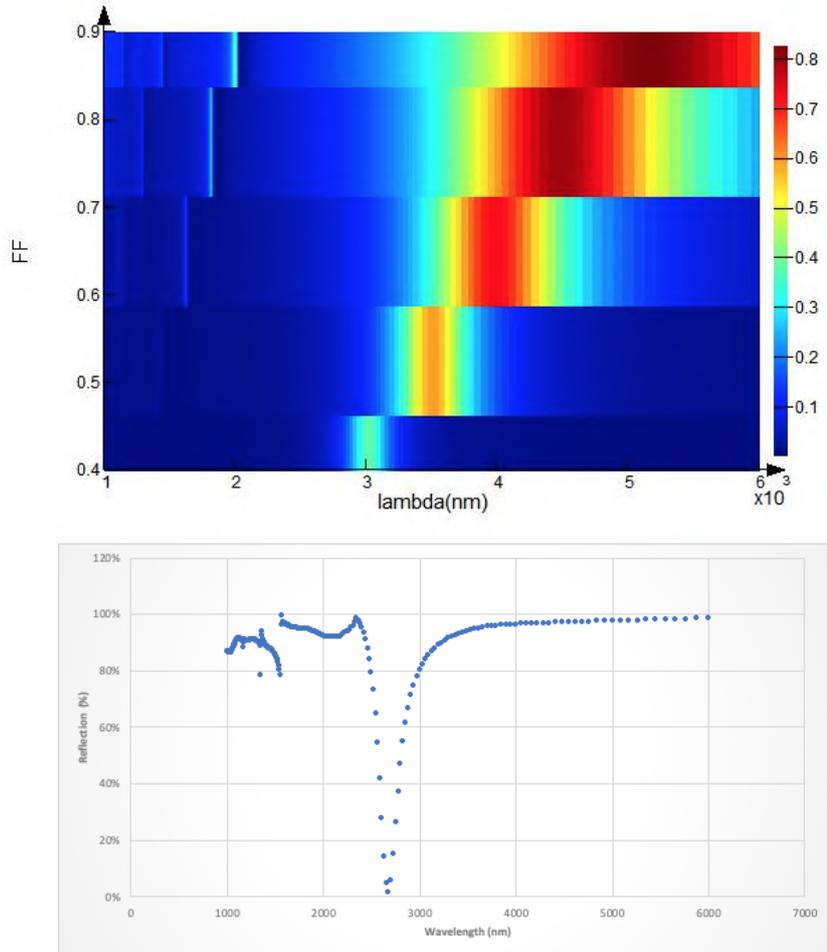

**Supplementary Figure S5**

*(Top) Effect of GST film thickness on PNA device performance. The contraction of the GST shows no negative effects on the device resonance. Transmission is increased, as expected, due to the reduced absorption length in the thinner GST film. (middle) Fill factor (FF), defined here as the ratio of hole diameter to array period, vs. transmission for the proposed device. As hole diameter increases the transmission also increases (as expected) however, the Q factor is rapidly deteriorated. A ratio of 0.4 was chosen for the device presented in this manuscript. (bottom) Plot of the reflectance spectrum of the presented PNA device showing perfect reflection away from resonance. This confirms a lack of absorption in the resonance.*

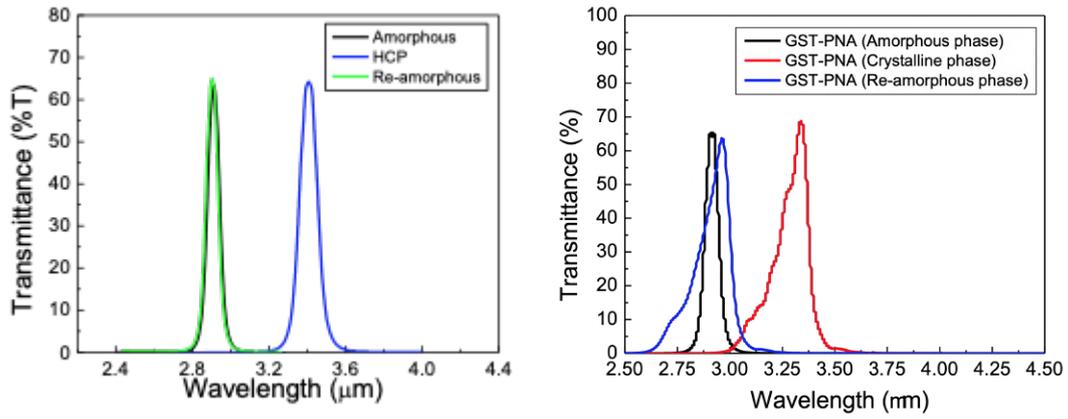

**Supplementary Figure S6**
*Measured transmission characteristics of the GST PNA device both with (left) and without (right) the ZnS:SiO$_2$ capping layer (thickness < 5 nm). Without the capping layer, the resonance shape is dramatically and irreversibly shifted as a result of surface oxidation and volatization of the GST film. The capping layer mitigates these effects and maintains the resonance quality across switching cycles.*

# Supplementary Note 3:
## Laser Characterization Setup

The laser used is a Quantel Evergreen (PN: EHP2715111) Nd:YAG operating at 532 nm and has 340mJ max energy output. The laser has a flat-top output shape rather than a Gaussian profile. The beam shape characteristics can be measured using a camera. However, to calculate energy density, both energy and beam diameter is needed. The output beam of the laser is 9.82 mm. However, the beam is apertured to 7 mm in order to ensure a uniform spatial profile. The beam is then directed towards the detector (Instrument – PN: FieldMaxII TOP, Accessory – PN: J-50MB-YAG, Coherent). Using a wedge window 99% of the beam energy is deflected into the energy detector. Energy density can then calculated using the energy from this detector and the diameter of the aperture. The other 1% of the split beam is directed into a CCD detector in order to characterize the beam profile (Newport LBP2-HR-VIS).

# Supplementary Note 4:
## IR Imaging Demonstration

In order to demonstrate the utility of the presented filters in an imaging system, the setup below was used. The components of the system are described in Figure S7. Although the system in this section is used passively, we describe in Supplementary Note 5 a method for implementing real-time tuning of the presented imaging system. In order to evaluate the performance of the filters in the imaging system, a comparison between measured data (in terms of counts at the detector) and expected values based on Planck's Law calculations (Figure S8). The measured data is in strong agreement with expected values.

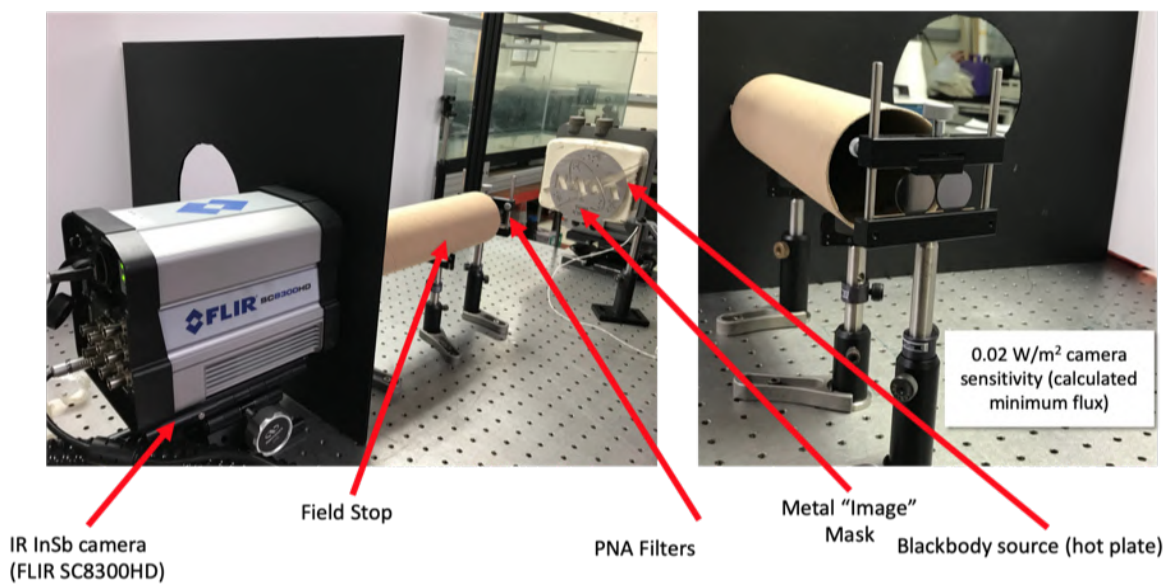

**Supplementary Figure S7: Image of MWIR Imaging Setup**

*Photograph of the imaging setup used in the manuscript. An InSb MWIR camera is used in conjunction with a field stop to reduce the field of view to include only the desired thermal object (in order to reduce noise in the measurements). Two GST PNA filters are placed in front of the field stop in order to be imaged simultaneously. A hotplate is used as a blackbody radiator, and an Al plate bearing the NASA logo is used as an image mask.*

| Blackbody (BB) Temp | Center λ₀ | ½ Power -λ₀ | ½ Power +λ₀ | HPBW | BB Emissive Power @ λ₀ | Filter Transmittance @ λ₀ | Theory | Imager | Measured |
|---|---|---|---|---|---|---|---|---|---|
| °C | µm | µm | µm | µm | W/m² | W/m² | % | Counts | % |
| 213.0 | 2.910 | 2.873 | 2.947 | 0.074 | 3.3 | 2.2 | | 2668.1 | |
| °K | 3.000 | 2.963 | 3.037 | 0.074 | 3.9 | 2.5 | 13.8% | 3098.2 | 16.1% |
| 486.15 | 3.110 | 3.073 | 3.147 | 0.074 | 4.6 | 3.0 | 18.4% | 3662.2 | 18.2% |
| | 3.340 | 3.303 | 3.377 | 0.074 | 6.2 | 4.0 | 34.9% | 5034.3 | 37.4% |
| Transmittance | 3.410 | 3.373 | 3.447 | 0.074 | 6.7 | 4.4 | 8.1% | 5433.9 | 7.9% |
| 65% | 2.910 | 2.873 | 2.947 | 0.074 | 3.3 | 2.2 | | 2668.1 | |

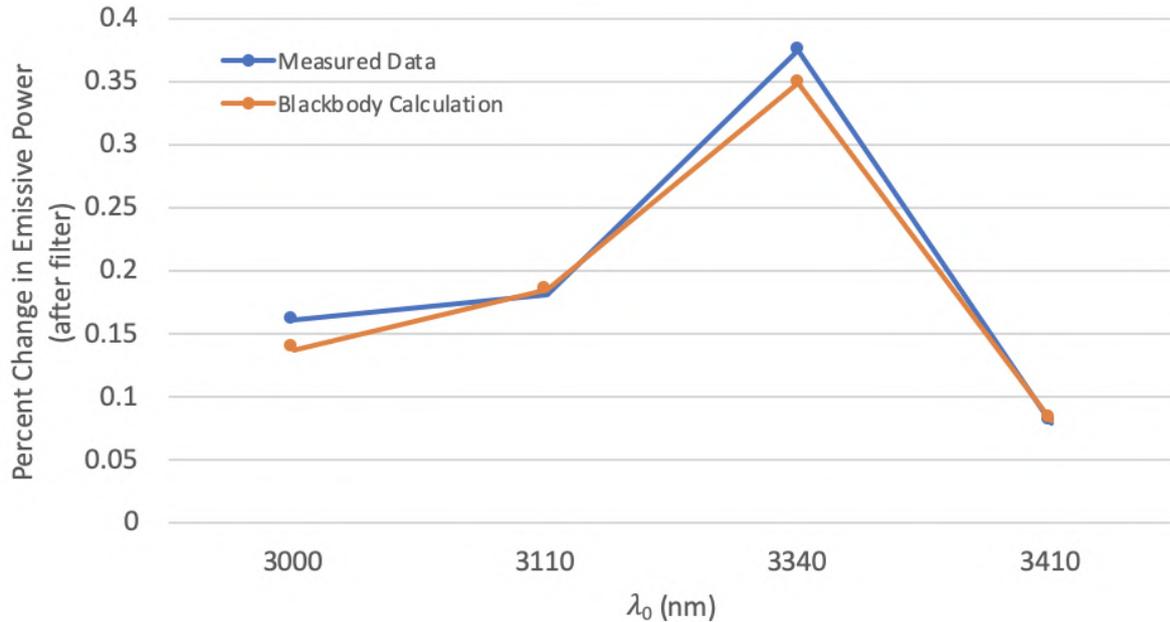

**Supplementary Figure S8: % comparison - BB / exp.**
*Comparison of the measured spectral irradiance vs. the expected values (after the PNA filter). The plot shows the % change in emissive power for both cases. As can be seen in the plot, the lines are largely overlapping, representing a very low experimental error.*

# Supplementary Note 5:
## Active "Read/Write" System Concept: Laboratory / Real World Systems

Although a laser-based actively tunable imaging system is sufficient for a laboratory environment, for a real-world application a more compact system must be realized. In order to achieve a compact system, a flashlamp can be employed consisting of a number of small, high powered diode sources with short pulse widths (as are commercially available - www.hamamatsu.com/us/en/product/lasers/semiconductor-lasers/plds/index.html). A small flashlamp can also be used, as these are typically quite compact and have large output powers (www.coherent.com/assets/pdf/PulseLife-G-Stack-DataSheet.pdf). Lastly, high-powered, ultra-compact lasers are also commercially available, and could work in a real-world system if integrated similar to a common (albeit higher-power) DVD/CD writer. Figure S9 shows an initial active imaging system, to be implemented in the next steps of this work. Following demonstration, the flashlamp system will be implemented followed by all electrical tuning.

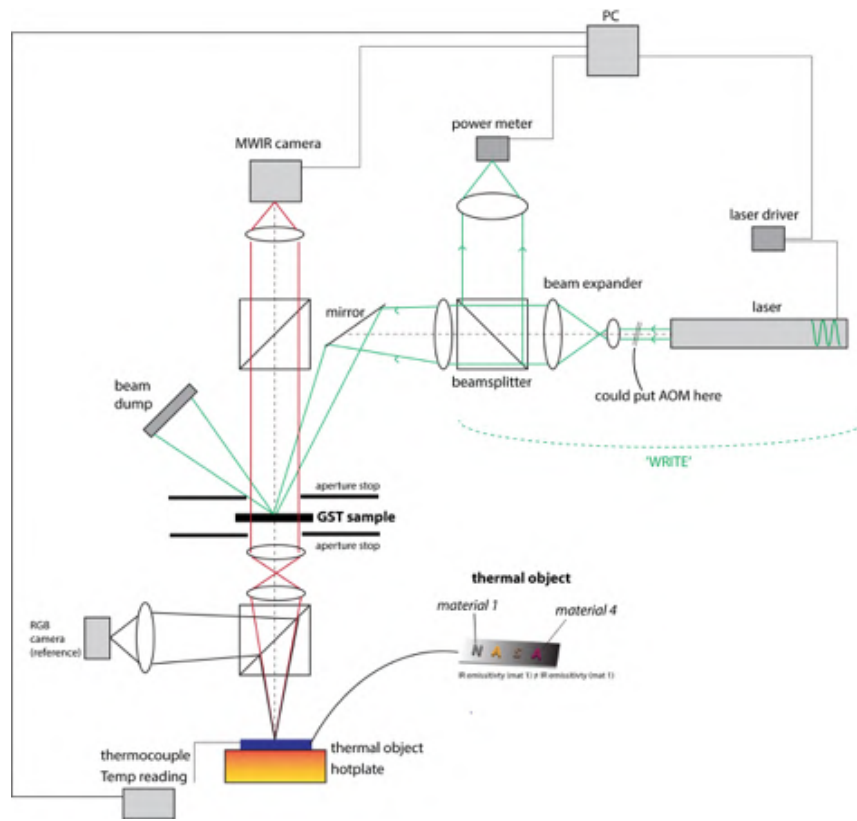

**Supplementary Figure S9: Laboratory-scale Active Imaging System**
*Schematic of a laboratory-based actively tunable GST PNA thermal imaging system capable of real-time active tuning. The setup is based on the switching setup presented in the main manuscript. In addition to the main "write/rewrite" system, which is now incident on the GST at a non-normal angle, a thermal source is placed behind the GST filter and an IR camera placed in front. In this configuration, simultaneous "read" and "write" functions can be readily realized.*